\documentclass[preprint2]{aastex}

\shorttitle{A Porous, Layered Heliopause}            
\shortauthors{Swisdak et al.}

\usepackage{amsmath}
\usepackage{amssymb}
\usepackage{bm}
\usepackage{url}
\usepackage{lineno}

\begin{document}

\title{A Porous, Layered Heliopause}

\author{M.~Swisdak\altaffilmark{1}, J.~F.~Drake\altaffilmark{2},
  M.~Opher\altaffilmark{3}}

\altaffiltext{1}{Institute for Research in Electronics
  and Applied Physics, University of Maryland, College Park, MD 20742,
  USA; swisdak@umd.edu}

\altaffiltext{2}{Department of Physics and the Institute for Physical
  Science and Technology and the Institute for Research in Electronics
  and Applied Physics, University of Maryland, College Park, MD 20742,
  USA; drake@umd.edu}

\altaffiltext{3}{Department of Astronomy, Boston University, 725
Commonwealth Avenue, Boston, MA 02215, USA;
  mopher@bu.edu}

\begin{abstract}

The picture of the heliopause (HP) -- the boundary between the domains
of the sun and the local interstellar medium (LISM) -- as a pristine
interface with a large rotation in the magnetic field fails to
describe recent Voyager 1 (V1) spacecraft data. Magnetohydrodynamic
(MHD) simulations of the global heliosphere reveal that the rotation
angle of the magnetic field across the HP at V1 is
small. Particle-in-cell simulations, based on cuts through the MHD
model at the location of V1, suggest that the sectored region of the
heliosheath (HS) produces large-scale magnetic islands that reconnect
with the interstellar magnetic field and mix LISM and HS plasma. Cuts
across the simulation data reveal multiple, anti-correlated jumps in
the number densities of LISM and HS particles at the magnetic
separatrices of the islands, similar to those observed by V1. A model
is presented, based on both the observations and simulation data, of
the HP as a porous, multi-layered structure threaded by magnetic
fields. This model further suggests that, contrary to the conclusions
of recent papers, V1 has already crossed the HP.

\end{abstract}


\section{INTRODUCTION}\label{intro}

The Voyager 1 and 2 spacecraft have been mapping the structure of the
outer heliosphere as they leave the solar system. In 2005, V1 crossed
the termination shock \citep{stone05a,burlaga05a,decker05a}, where the
supersonic solar wind becomes subsonic, and has since been traversing
the HS. The HP, whose location and structure are unknown, separates
the magnetic field and plasma associated with the sun from that of the
LISM \citep{parker63a,baranov79a}.  The magnetic field in the HS has
remained dominantly in the azimuthal (east-west) direction given by
the Parker spiral but could rotate and acquire measurable north-south
and radial components upon crossing the HP. In ideal (non-dissipative)
models of the heliosphere, the local magnetic field is transverse to
the boundary and the HP is a tangential discontinuity
\citep{parker63a,baranov79a}. However, whether the HP is a smooth
interface, or breaks up due to instabilities, has been the subject of
substantial discussion in the literature
\citep{fahr86a,baranov92a,liewer96a,zank96b,swisdak10a}. The structure
of the HP, and in particular whether the boundary is porous to some
classes of particles, is of great importance because of its impact on
the transport of energetic particles into and out of the heliosphere.

Beginning on day 210 of 2012, the V1 spacecraft measured a series of
dropouts in the intensities of energetic particles produced in the
heliosphere: the Anomalous Cosmic Rays (ACRs) and the lower-energy
Termination Shock Particles (TSPs)
\citep{webber13a,stone13a,krimigis13a}. Simultaneous with the dropouts
were abrupt increases in the Galactic Cosmic Ray (GCR) electrons and
protons and an increase in the magnetic field intensity
\citep{burlaga13a}. Finally, on around day 238, the
heliospheric-produced particles dropped to noise levels and the GCRs
underwent a final increase.  Both have since exhibited no significant
variations, which suggests that V1 crossed the HP, with the repeated
dropouts and increases perhaps due to radial fluctuations caused by
changes in the solar wind dynamic pressure. However, during this time
the direction of the magnetic field remained dominantly azimuthal
\citep{burlaga13a}, consistent with the spacecraft remaining in the
HS. While MHD models of the heliosphere suggested that the rotation of
the magnetic field across the HP at the location of V1 would be small
\citep{opher09a,opher09b}, the lack of any significant change in the magnetic
field direction across the final transition on day 238 suggested that
V1 remained within the magnetic domain of the HS.

We present here a model of the magnetic structure of the HP at V1's
location that produces particle and magnetic signatures consistent
with the observations.  By pairing a global MHD simulation with a
local PIC simulation, we show that magnetic reconnection can produce a
complex, nested set of magnetic islands at the HP. Tongues of LISM
plasma penetrate into the HS along reconnected field lines. These
tongues correspond to local depletions of the HS plasma and
enhancements in the magnetic pressure. A key result of the simulations
is that sharp anti-correlated jumps in the HS and LISM number density
can occur across the separatrices emanating from reconnection sites
while the magnetic field undergoes essentially no rotation. Such
behavior undercuts the primary argument suggesting that V1 has not
crossed the HP -- that no field rotation was seen on day 238 where the
final drop in ACRs was measured \citep{burlaga13a}. We therefore
suggest that V1 actually crossed the HP on day 209, the time of the
last reversal in the azimuthal magnetic field $B_T$, and that the
steady values of the normal $B_N\sim 0.12\text{ nT}$ and $B_T\sim
-0.40\text{ nT}$ fields since that time are the draped interstellar field just
outside of the HP.

\section{MHD Simulation}\label{MHD}
To establish local conditions at the HP, we first explore the
heliosphere's large-scale structure with a global MHD simulation that
includes both neutral and ionized components (and both thermal and
pick-up ions in the solar wind) \citep{zieger13a}.  The LISM field,
$B_{\text{ISM}}$, has a magnitude of $0.44\text{ nT}$ and a direction
defined by $\alpha_{Bv}=15.9^{\circ}$ and $\beta_{Bv}=51.5^{\circ}$,
where $\alpha_{Bv}$ and $\beta_{Bv}$ are the angle between
$B_{\text{ISM}}$ and the velocity of the interstellar wind
$v_{\text{ISM}}$ and the angle between the
$B_{\text{ISM}}$-$v_{\text{ISM}}$ plane and the solar equator
\citep[for further discussion of these choices see][]{opher09a}. The
$Z$-axis is along the solar rotation axis and the $X$-axis is chosen
so that $v_{\text{ISM}}$ lies in the $X$-$Z$ plane.  The MHD
simulation did not include the sector zone (where the Parker spiral
field periodically reverses polarity due to the tilt between the solar
magnetic and rotation axes) since this leads to field reversals that
cannot be numerically resolved upstream of the HP and therefore
produces incorrect values of $B=|\mathbf{B}|$
\citep{opher11a,borovikov11a}. The solar field polarity corresponds
to solar cycle 24, with the azimuthal angle $\lambda$ (between the
radial and $T$ directions in heliospheric coordinates) $90^{\circ}$ in
the north and $270^{\circ}$ in the south.

\begin{figure}
 \centering \includegraphics[width=20pc]{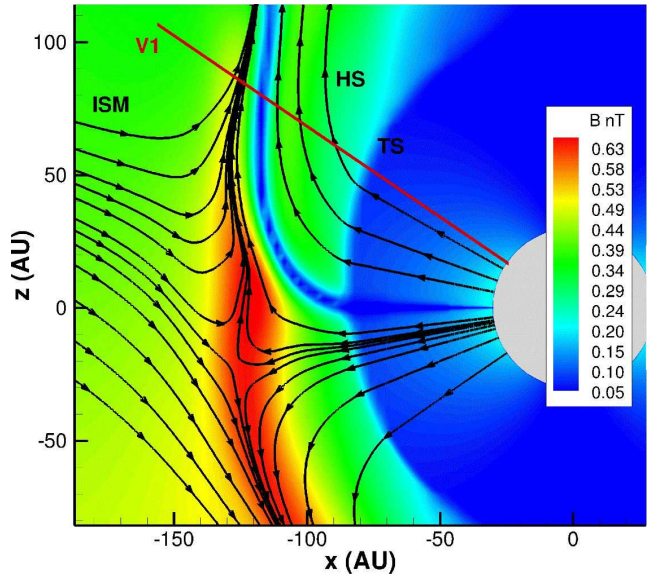}
 \caption{\label{MHDfig}A meridional cut from the global MHD
 simulation showing the magnetic field amplitude $B$ (background), the
 flow streamlines (solid curves with arrows) and the V1 trajectory
 (red). The HP is where the flows from the LISM and the HS meet. The
 blue line in the HS is the heliospheric current sheet.}
\end{figure}

In Fig.~\ref{MHDfig}, $B$ from the simulation reveals the solar wind
compression at the termination shock, the downstream HS, and the
HP. Profiles (solid curves in Fig.~\ref{cutsfig}) along the V1
trajectory of the pick-up ($n_{\text{pui}}$) and thermal
($n_{\text{th}}$) ion densities and the azimuthal ($B_T$) and normal
($B_N$) magnetic fields near the HP are inputs for the PIC
simulations.  $n_{\text{pui}}$ decreases from $\approx 7\times
10^{-4}/\text{cm}^3$ in the HS to zero in the LISM while
$n_{\text{th}}$ rises from $0.003/\text{cm}^3$ to $\approx
0.08/\text{cm}^3$.  $B_N$ (Fig.~\ref{cutsfig}C) is small at V1's
latitude in the LISM. $B_T$ flips direction across the HP, but remains
the dominant component on both sides of the boundary
(Fig.~\ref{cutsfig}D). The polar angle $\delta$ (the
angle between $B_N$ and the equatorial field) in the simulations approaches
$14^{\circ}$ just outside of the HP, which is consistent with the steady values
seen in the V1 data.

The MHD simulation does not match Voyager observations in several
respects --- the sign of the HS azimuthal magnetic field
orientation, the strength of the flows in the HS and the
characteristic scale length of the HP --- none of which is essential
for calculating initial conditions for the PIC simulations. First,
because V1 continued to measure sector boundaries in the HS during
2012, and therefore probably remained in the sector zone, the sign of
$B_T$ in the HS in the MHD model is irrelevant since a ``correct''
model should include the reversals associated with the sector region.
Second, in contrast to the simulation, indirect measurements by the V1
LECP instrument indicate little to no normal flow in the HS
\citep{decker12b}.  No published global model has explained the
observed flows, although simulations that include the sectored field
\citep[e.g.,][]{opher12a} are closer to the observations than those presented
here \citep[see also][for an alternative
explanation]{pogorelov12a}. Finally, since the MHD model does not
include the physics necessary to describe the structure of the HP, the
scale length of this transition is not physical, but is instead a
numerical artifact.  On the other hand, what is essential for input
into the PIC simulations is the strength of the HS field and the
strength and orientation of the field in the LISM.

\section{PIC Simulations}\label{PIC}  
\begin{figure}
 \centering \includegraphics[width=20pc]{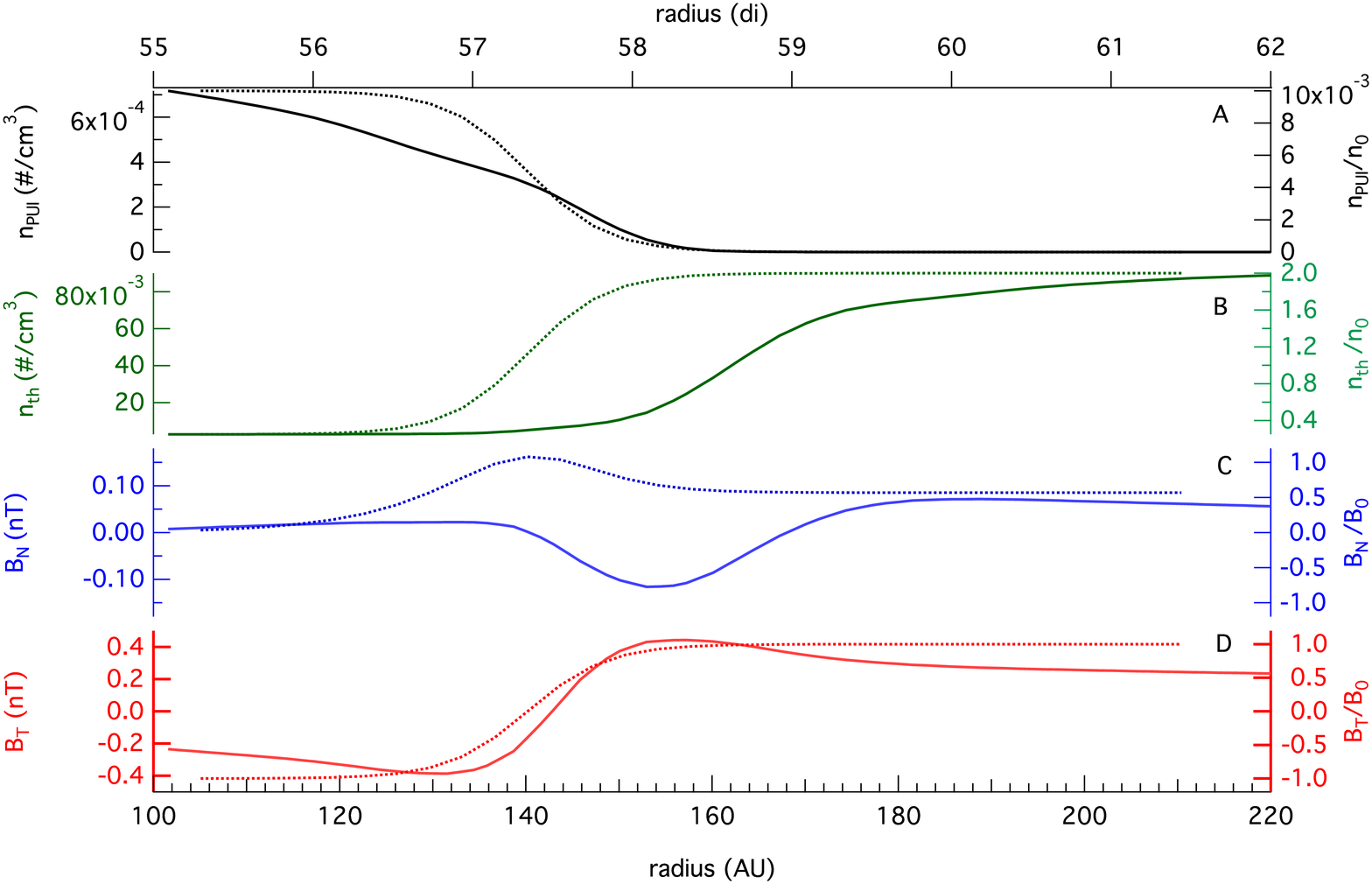}
 \caption{\label{cutsfig} Cuts of various parameters along the V1
 trajectory near the HP from the MHD model (solid with left scale) and
 as initial conditions for the PIC model (dotted with right
 scale). Shown are in (A) the pick-up ion density, in (B) the thermal
 ion density, in (C) $B_T$ and in (D) $B_N$. Note that the scales on
 the right and left differ.}
\end{figure}

The initial profiles of the magnetic field, density, and temperature
for the 2-D PIC simulations (dotted lines in Fig.~\ref{cutsfig};
right-hand scale) were constructed with input from the MHD profiles
although, in keeping with the Voyager 1 observations, there are no
initial flows. The PIC code is written in normalized units based on a
field strength $B_0$ and density $n_0$ (lengths normalized to the ion
inertial length $d_i =c/\omega_{pi}$, with $\omega_{pi}$ the ion
plasma frequency, times to the ion cyclotron time $\Omega_{i0}^{-1}$
and velocities to the Alfv\'en speed $c_{A0}$). In the HS, thermal
($n_{\text{th}}=0.25n_0$, $T_{\text{th}}=0.25 m_ic_{A0}^2$) and
pick-up ions ($n_{\text{pui}}=0.01n_0$,
$T_{\text{pui}}=15.0m_ic_{A0}^2$) were included as independent species
while the LISM only included a thermal component ($n_{\text{th}} =
2.0n_0$, $T_{\text{th}}=0.2m_ic_{A0}^2$).  The simulations were
performed in a domain with dimensions $(L_T, L_R)= (409.6d_i,
204.8d_i)$. The ion-to-electron mass ratio was 25 and the velocity of
light was 15$c_{A0}$.  Not shown in Fig.~\ref{cutsfig} are the three
current sheets, of initial half-width $0.5 d_i$, that produce the
sectored HS field.  This scale reflects satellite measurements at the
Earth's magnetopause that such boundaries collapse to kinetic scales
\citep{sonnerup81a}. Pressure balance across each reversal is achieved
by adjusting the out-of-plane component $B_N$ \citep{smith01a}.

For HS-appropriate values, $n = 10^{-2} \text{
cm}^{-3}$ and $B = 0.3 \text{ nT}$, $d_i \approx 2\times 10^{-5}
\text{ AU}$, $\Omega_{ci}^{-1} \approx 30 \text{ s}$, and $c_{A}
\approx 100 \text{ km/s}$.  Resolving kinetic scales forces the
simulation domain to be much smaller than the actual system.  Despite
this limitation, the important physical processes can still be
understood by appropriately scaling the results \citep{schoeffler12b}.

\begin{figure}
 \centering \includegraphics[width=20pc]{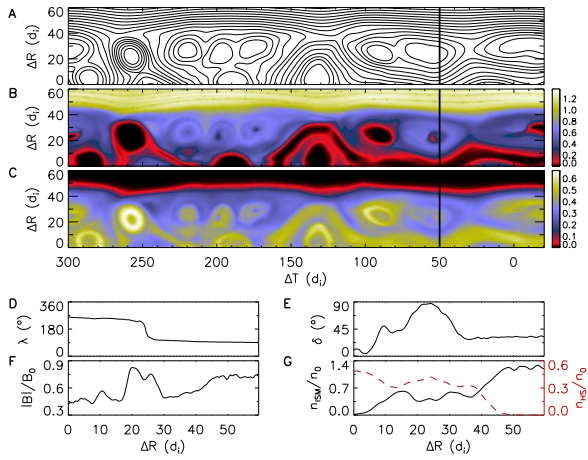}
 \caption{\label{PICfig}The structure of the HP and adjacent LISM
 and HS at late time. In the $R$/$T$ plane in (A) the
 magnetic field lines and in (B) and (C) the number density
 $n_{\text{LISM}}$ ($n_{\text{HS}}$) of particles that were originally
 in the LISM (HS), respectively. Panels (D)-(G) are cuts along the
 vertical line in panels (A)-(C).  In (D) the azimuthal angle
 $\lambda$ is the angle of $\mathbf{B}$ in the $R-T$ plane with
 respect to the $R$ direction.  In (E) the polar angle $\delta$ is the
 angle between $\mathbf{B}$ and the $R-T$ plane.  In (F), the
 magnitude of B and, in (G), the number density $n_{\text{LISM}}$
 (solid) and the number density $n_{\text{HS}}$ (dashed red).}
\end{figure}

The simulations are evolved with no initially imposed
perturbations. Because of the lower density, which leads to a locally
higher $c_A$ and effectively thinner current sheets (when normalized
to the local $d_i$), magnetic reconnection first starts in the
sectored HS.  Small magnetic islands grow on individual current layers
in the HS and merge to become larger islands until they are comparable
in size to the sector spacing (Fig.~\ref{PICfig}A). A chain of small
islands grows at the HP. These merge, forming larger islands, and are
compressed by HS islands pushing against the HP
(Fig.~\ref{PICfig}A). By late time, the HS magnetic field has
reconnected with that of the LISM, forming a complex, nested chain of
islands (Fig.~\ref{PICfig}A) at the HP with sizes comparable to the
original sector spacing. Along a cut from the HS to the LISM (dark
line in Fig.~\ref{PICfig}A) the HP is at $\Delta R/d_i\sim 25$ where
$B_T$ reverses sign (Fig.~\ref{PICfig}D). The rotation to the LISM
field direction is complete by $\Delta R/d_i\sim 35$ after which
$\delta$ and $\lambda$ are nearly constant.



We can independently track every particle in the PIC model and
therefore can explore the mixing of the LISM and HS plasmas.  The
overall result is a highly structured distribution in the densities of
the LISM ($n_{\text{LISM}}$) and HS ($n_{\text{HS}}$), with each
experiencing sharp jumps across the separatrices bounding the outflows
ejected from reconnection sites (Fig.~\ref{PICfig}A-C). Particles
initially in the LISM continue to dominate the density on the
un-reconnected LISM field lines, have mixed with HS particles in the
nested islands formed from HS-LISM reconnection, and are largely
excluded from islands formed from reconnection of the HS sectored
field (Fig.~\ref{PICfig}B). Particles initially in the HS dominate
islands resulting from reconnection of the sectored field, are mixed
with LISM particles in the HP islands, and are nearly excluded from
un-reconnected regions of the LISM (Fig.~\ref{PICfig}C).

Radial cuts through the simulation reveal that the increases and
decreases in $n_{\text{LISM}}$ and $n_{\text{HS}}$ are typically
anti-correlated (Fig.~\ref{PICfig}G). Moving from a pure HS magnetic
island into an island or outflow jet where LISM and HS plasma has
mixed reduces $n_{\text{HS}}$ and increases $n_{\text{LISM}}$. Along the 
cut the first drop in $n_{\text{HS}}$ occurs downstream of a magnetic
separatrix, where HS particles have an open path to the LISM along
open field lines ($\Delta R/d_i \sim 8$ in
Fig.~\ref{PICfig}G). Similar behavior has been documented in satellite
measurements at the Earth's magnetopause \citep{sonnerup81a} and
echoes V1's observations of the anti-correlated variations in the
fluxes of ACRs/TSPs and galactic electrons/GCRs
\citep{stone13a,krimigis13a,webber13a}. Most important, the cuts
further reveal that, when crossing the last magnetic separatrix on the
LISM side of the HP before finally entering the pristine LISM plasma,
the sharp decrease in $n_{\text{HS}}$ and increase in
$n_{\text{LISM}}$ (in the interval $\Delta R/d_i = 38-50$ in
Fig.~\ref{PICfig}G) occur over an interval where there is no
directional change in the magnetic field (Fig.~\ref{PICfig}D-E). The
absence of a directional change in $\mathbf{B}$ at locations with
strong variations in the particle densities is consistent with one of
the most significant of the V1 observations \citep{burlaga13a}. The
fact that, in our simulation, this occurs on the LISM side of the HP
therefore suggests that it may be incorrect to conclude that V1 has
not crossed the HP.

The simulation cuts also reveal that local decreases in the HS density
typically correspond to increases in the local magnetic field
(Figs.~\ref{PICfig}F-G). The total pressure across the HP is
balanced. While the dominant pressure in the HS is from the plasma,
the dominant pressure in the LISM is magnetic. Thus, when reconnection
opens a path for HS plasma to escape into the LISM and mix with the
lower-pressure LISM plasma, there is nothing to balance the total
pressure and so the region compresses to increase the magnetic field
amplitude. This behavior is primarily seen at separatrix crossings
remote from where reconnection locally reduces the magnetic field
strength (the interval $\Delta R/d_i=38-50$ in Figs.~\ref{PICfig}F-G). In
the V1 data, the magnetic field strength is also observed to increase
where the local flux of HS plasma decreases
\citep{stone13a,krimigis13a,webber13a,burlaga13a}.

\begin{figure}
 \centering \includegraphics[width=20pc]{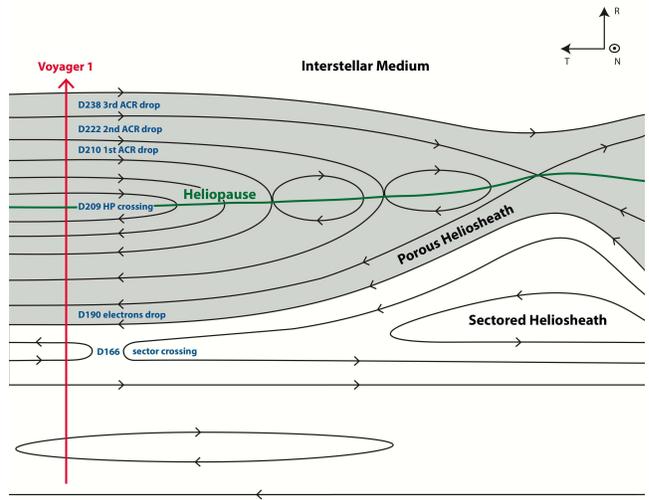}
 \caption{\label{cartoonfig}A schematic, based on the results of our
 simulations, of the inferred magnetic structure of the HP during the
 time when V1 documented strong variations in the HS and LISM
 particles.  The times corresponding to several of the Voyager events
 are marked by the days of 2012 on which they occurred.}
\end{figure}

Thus, based on our simulations, we suggest that the V1 observations of
simultaneous drops (increases) in HS (LISM) particle fluxes occur at a
series of separatrix crossings outside of the HP that are associated
with a nested set of magnetic islands that form at the HP
(Fig.~\ref{cartoonfig}). At such crossings the magnetic field
direction does not change significantly, while, as seen in the
simulation data, particle fluxes can change sharply. Three active
reconnection sites at the HP, and associated separatrices with two
nested islands, are sufficient to explain the sequence of Voyager
events. On day 166 the spacecraft crossed a current layer, on day 190
the flux of HS electrons began dropping, on day 209 another current
layer was crossed and on days 210, 222 and 238 three successive drops
(increases) in the HS (LISM) particle fluxes occurred. The day 190
drop in the HS electrons suggests that after this time the magnetic
field was no longer laminar so that these electrons, with their small
Larmor radii and large velocities, could leak into the LISM. 

Islands and x-lines flowing away from an active x-line (e.g., the
rightmost x-line in Fig.~\ref{cartoonfig}) correspond to reconnection
sites that developed earlier in time. The separatrix field lines
connect to x-lines, which act as bottlenecks to particle transport
across the HP. There are two reasons for this. First, at the x-line
the magnetic field turns into the $N$ direction since the $R$ and $T$
magnetic field components are zero. Thus, the x-line halts the
field-aligned streaming of particles across the HP. Second, to the
extent that $B_N$ is weak compared with $B_T$ energetic particles can
scatter near x-lines, which further limits transport across the HP. In
contrast, downstream of separatrices (to the left in
Fig.~\ref{cartoonfig}) particles can freely stream across the
HP. Thus, the day 210 drop in ACRs (rise in GCRs) occurred at the
separatrix corresponding to the left-most x-line of
Fig.~\ref{cartoonfig} which blocked the transport of ACRs (GCRs)
across the HP. The ACR (GCR) intensity rose (dropped) as the
spacecraft crossed field lines that formed an open corridor across the
HP to the left of the middle x-line in Fig.~\ref{cartoonfig}.  In this
region the flow of GCRs into the HS acts as a sink for the intensity
of these particles. The second ACR drop on day 222 and subsequent
recovery is similar. The final drop of the HS particle fluxes on day
238 occurred at the separatrix of the right-most x-line in
Fig.~\ref{cartoonfig}. LECP measurements of ACR anisotropies
\citep{krimigis13a} show that particles propagating parallel to the
magnetic field dropped more rapidly than those with perpendicular
pitch angles on day 238.  Such behavior is consistent with our
schematic -- in a weakly stochastic field parallel moving particles
will more quickly escape in the LISM.

An inconsistency between our PIC simulations and the observations
concerns the spatial region where sharp jumps in the ACRs and GCRs
take place. In the simulations the anti-correlated jumps occur on both
sides of the HP but if our cartoon is correct, they occur only on the
LISM side in the V1 data. The contradiction is possibly because the
simulations are in a 2-D system. A magnetic field in a real 3-D system
will likely be at least mildly stochastic so that wandering field
lines will smooth the variability of ACR and GCR intensities far from
the HP boundary. More challenging 3-D simulations will be required to
explore this issue.

A second issue is the angle $\delta$ of the magnetic field outside of
the HP, which for the present MHD simulation is
$14^{\circ}$. $B_{ISM}$ for this simulation was chosen to match
heliospheric asymmetries \citep{opher09a,opher09b}. Other MHD
simulations based on fitting the IBEX ribbon yield $\delta\sim
30^{\circ}$ \citep{pogorelov13a}. Such values are considerably smaller
than those implied by cartoons in recent publications
\citep{burlaga13a}. In any case our conclusion that significant
variations in the density of the ACRs and GCRs can occur in regions
with essentially no variation in the field line orientation is not
sensitive to the value of $\delta$ in the LISM. Any rotation in the
magnetic field outside of the HP propagates at the local Alfv\'en
speed, which is well below the velocities of the particles of
interest. Thus, outside of the HP separatrices should retain their
original LISM orientation in locations where there are significant
variations in particle intensity.

Combining a typical aspect ratio of reconnection-produced islands
(0.1), the typical time between dropouts (10 days), and the speed of
Voyager 1 with respect to the HS plasma ($\approx 20\text{ km/s}$),
yields the approximate size of the islands in Fig.~\ref{cartoonfig} as
1 AU. This roughly equals the sector spacing downstream of the TS and
is consistent with previous simulation-derived estimates of magnetic
islands in the HS \citep{schoeffler12b}.

If the schematic with nested magnetic islands (Fig.~\ref{cartoonfig})
is correct, the dropouts in the HS particle fluxes occurred on the
LISM side of the HP on field lines that had a LISM source. Thus,
according to this picture V1 crossed the HP on day 208 and has been
crossing LISM fields since that time.  Our results thus suggest that
$B_T$ in the LISM is negative (i.e., has a polarity of $270^{\circ}$).


\begin{acknowledgments}

The authors acknowledge the support of NSF grant AGS-1202330 to the
University of Maryland, and NSF grant ATM-0747654 and NASA grant
NNX07AH20G to Boston University. The PIC simulations were performed at
the National Energy Research Scientific Computing Center and the MHD
simulations at the Ames NASA Supercomputer Center. We acknowledge
fruitful discussions with L. F. Burlaga, R. B. Decker, M. E. Hill and
E. C. Stone on the Voyager observations. This research benefited
greatly from discussions held at the meetings of the Heliopause
International Team at the International Space Science Institute in
Bern, Switzerland.

\end{acknowledgments}



\newpage

\end{document}